\documentclass[prd, twocolumn, nofootinbib, floatfix]{revtex4-1}

\usepackage{amsmath}
\usepackage{graphicx}
\usepackage{dcolumn}
\usepackage{bm}
\usepackage{epsfig}
\usepackage{amssymb,latexsym,mathrsfs}
\usepackage{graphicx}
\usepackage{color}
\usepackage{hyperref}
\usepackage{subfigure}
\usepackage[normalem]{ulem}

\hypersetup{
    colorlinks=true,
    linkcolor=red,
    citecolor=blue,
} 

\newcommand{\be}{\begin{equation}}
\newcommand{\ee}{\end{equation}}
\newcommand{\bs}{\begin{split}} 
\newcommand{\bea}{\begin{eqnarray}}
\newcommand{\eea}{\end{eqnarray}}

\newcommand{\al}{\alpha}

\renewcommand{\d}[1]{\ensuremath{\operatorname{d}\!{#1}}}
 
\newcommand{\bo}{\raise-1mm\hbox{\Large$\Box$}}

\newcommand{\w}{\omega}

\begin{document}

\title{Slicing the Vacuum: New Accelerating Mirror Solutions of the Dynamical Casimir Effect} 
\author{Michael R.R. Good${}^{1,2}$}
\author{Eric V.\ Linder${}^{2,3}$} 
\affiliation{${}^1$Physics Department, School of Science and Technology, Nazarbayev University, Astana, 
Kazakhstan\\
${}^2$Energetic Cosmos Laboratory, Nazarbayev University, Astana, 
Kazakhstan\\ 
${}^3$Berkeley Center for Cosmological Physics \& Berkeley Lab, 
University of California, Berkeley, CA 94720, USA} 

\begin{abstract}
Radiation from accelerating mirrors in a Minkowski spacetime 
provides insights into the nature of horizons, black holes, and 
entanglement entropy. We introduce new, simple, symmetric and analytic moving mirror solutions and study their particle, energy, and entropy production. This includes 
an asymptotically static case with finite emission that is the black hole analog of complete evaporation.  The total energy, total entropy, total particles, and spectrum are the same on both sides of the mirror. 
We also study its asymptotically inertial, drifting analog (which gives a black hole remnant) to explore 
differences in finite and infinite production. 
\end{abstract} 

\date{\today} 

\maketitle

\section{Introduction} 
Moving mirrors are the most elementary examples of the dynamical Casimir effect \cite{Moore:1970}.  
The study of moving mirrors as analog systems for investigating the 
nature of horizons and associated issues for black holes particle creation \cite{Hawking:1974sw}, including 
information paradoxes, has a long tradition \cite{DeWitt:1975ys, Davies:1976hi, Davies:1977yv}. 
Particularly useful are analytic solutions where one can obtain a 
deeper intuition into the particle production, energy flux, entanglement entropy, and analogy with black hole 
spacetimes. 

We develop two new solutions that are asymptotically inertial.  In the far past and far future, one is
static (zero velocity), while the other is drifting (zero acceleration, coasting with constant velocity). They both have infinite travel (covering infinite coordinate distance) but are ultimately ``unmoved'' (the asymptotic past and future positions 
are identical, at infinite coordinate position). We relate these properties and symmetries to the particle production, energy emission, and 
entanglement characteristics. 

In addition we give a general prescription for regularizing asymptotically 
infinite acceleration solutions to finite acceleration or inertial results. 
This enables the generation of a new testbed of solutions from known ones. 

In Sec.~\ref{sec:method} we outline our general approach and expectations 
for the relation between trajectory properties and particle production. 
Specific solutions are treated in Sec.~\ref{sec:solutions}, including the 
particle spectra, while we discuss analogies with black hole spacetimes 
and information/entropy in Sec.~\ref{sec:blackhole}. We summarize and 
mention future work in Sec.~\ref{sec:concl}. The units are $k_B = \hbar = c = 1$.

\section{General Properties} \label{sec:method} 

Our explorations take place in a $1+1$ dimensional Minkowski spacetime, 
with a massless scalar field interacting with a moving mirror. The 
trajectory of the mirror, $z(t)$, can be related to the null coordinates, 
or ``shock functions''\cite{Good:2016atu}, and from any of these the Bogolyubov coefficients 
of the particle creation operators can be calculated, as well as the ensuing 
particle spectrum and total particle count. The shock functions also provide the energy flux and total energy. The general methodology is well discussed in the literature, e.g. \cite{Carlitz:1986nh, Good:2012cp, Good:2016oey}. 

Only a handful of analytic solutions for the particle production 
characteristics are known (see \cite{Good:2016atu} for a summary). Despite the benefit of consistency for finite total particle production there are only two such analytic Bogolyubov $\beta$ coefficient solutions that are previously known: the Walker-Davies 
\cite{walkerdavies} and Arctx \cite{Good:2013lca} cases. We seek to add a new, more simple 
solution, so as to better understand the general properties and variations 
that can relate the mirror trajectory to the particle production and 
horizon analogs. 

In looking for analytically tractable solutions we consider trajectories subject to two strong constraints: (1) time-reversal symmetry and (2) asymptotic inertial character.  The benefits of these constraints are two-fold: the symmetry ensures simplicity and self-duality, and the asymptotically inertial character insures the total emitted energy remains finite.  The limititation to motions involving only cases that are either asymptotically static, or asymptotically drifting, rather than those motions that have acceleration singularities or those motions that accelerate forever, avoids the pathology of infinite energy production.  
We choose the mirror velocity $\dot z(t)$ as the starting point for the exploration, since it is directly related 
to such properties and we can immediately ensure that whatever functional form is explored is bounded such that $|\dot z| < 1$, where the speed of light is set to one.

\subsection{Regularizing acceleration} \label{sec:reg} 

Mirror solutions that are asymptotically inertial are expected to 
have finite energy production (though not necessarily finite particle 
production). It is therefore of interest to devise a way of generating 
such non-divergent energy solutions from cases with constant or infinite asymptotic acceleration, 
i.e.\ regularizing the acceleration. One method, the multiplicative 
shift, has been discussed in \cite{Good:2016atu}, and we here introduce a 
second method, the additive shift. 

A mirror's acceleration is given by the rectilinear proper acceleration
\be 
\al=\gamma^3 \ddot z=\frac{\ddot z}{(1-\dot z^2)^{3/2}} \ , 
\ee 
where $\gamma$ is the Lorentz factor. With a multiplicative shift 
in velocity, $\dot z=v\dot z_0$, a new trajectory, $z(t)=vz_0(t)$, 
is generated from an old one $z_0(t)$. The resulting acceleration is 
\be 
\alpha=\frac{v\ddot z_0}{(1-v^2\dot z_0^2)^{3/2}} \ . 
\ee 
If at some time the asymptotic velocity $\dot z_0(t)=0$, then this just scales 
$\alpha=v\alpha_0$, so it cannot create an inertial state from a 
noninertial one and can't regularize an infinite acceleration. If the 
asymptotic velocity $\dot z_0(\infty)=\pm1$, i.e.\ the speed of light, 
then the multiplicative shift can create an 
inertial state $\alpha(\infty)=0$ as long as the original solution 
had $\ddot z_0(\infty)=0$. 

The additive shift method can regularize an infinite acceleration even when 
the velocity $\dot z_0(t)=0$. With an additive shift 
in velocity, $\dot z=\dot z_0-v$, a new trajectory $z(t)=z_0(t)-vt$ 
is generated from an old one $z_0(t)$. The resulting acceleration is 
\be
\alpha=\frac{\ddot z_0}{[1-(\dot z_0-v)^2]^{3/2}}=
\alpha_0\,\left[1+\frac{v(2\dot z_0-v)}{1-\dot z_0^2}\right]^{-3/2} \ . 
\ee 
If the asymptotic velocity $\dot z_0(\infty)=0$, then the asymptotic acceleration is
\be 
\alpha=\frac{\alpha_0}{(1-v^2)^{3/2}} 
\ee 
so $v=\pm1$ could potentially 
regularize infinite acceleration, either to an inertial state or to 
a finite acceleration. If the asymptotic velocity $\dot z_0(\infty)=\pm1$, 
then 
\be 
\alpha=\frac{\ddot z_0}{(\pm2v-v^2)^{3/2}} \ . 
\ee 
This can give an inertial asymptote if $\ddot z_0(\infty)=0$ (the 
only case for which the multiplicative shift works). 
Thus the additive shift has one more potential way of dealing with 
infinite acceleration than the multiplicative shift. 

The additive shift method works to deliver inertial solutions 
from infinite acceleration ones. For example, two known mirror solutions, the eternally thermal infinite energy (Carlitz-Willey \cite{Carlitz:1986nh}; see \cite{Good:2012cp} or \cite{Good:2013lca} for the $z(t)$ trajectory) and the asymptotically light speed with finite energy (Proex \cite{Good:2013lca}) trajectories are additive shifts of one another (up to a form factor of 2). 

As far as asymptotically inertial trajectories are concerned, asymptotically static cases have features of interest even beyond asymptotically 
drifting ones.  One such feature is that their particle production is expected to be finite.  This was demonstrated in the two previously existing asymptotically static, 
analytic solutions: $(1)$ the two-parameter Walker-Davies \cite{walkerdavies} $t(z)$ trajectory, and $(2)$ the one-parameter Arctx \cite{Good:2013lca} $z(t)$ trajectory. In this note, we introduce a new,  simpler, asymptotically static solution (two-parameter, $z(t)$ trajectory). 

\subsection{Power-power family} 

A ratio of power laws in time, with one less power in the numerator, 
for the mirror velocity will give an asymptotically static state. We 
call this the power-power family and for the velocity 
\be 
\dot z=vn\,\frac{|t/\tau|^{n-1}}{|t/\tau|^n+1} \ , 
\ee 
we can analytically write the mirror trajectory and acceleration as 
\bea 
z&=& v\tau\,\ln\left(\left|\frac{t}{\tau}\right|^n+1\right), \\ 
\al&=& \frac{vn}{\tau}\,(r^n+1)\,\frac{(n-1)r^{n-2}-r^{2(n-1)}}{[(r^n+1)^2-v^2 n^2r^{2(n-1)}]^{3/2}} \ , 
\eea 
where $r\equiv|t/\tau|$, $v$ is related to the maximum velocity by $\dot z_{\rm max}=v(n-1)^{(n-1)/n}$, and 
$\tau$ is a characteristic time scale. The inverse of $\tau$ is proportional to the acceleration parameter $\kappa$, and we can choose $\kappa \tau \equiv 1$. In the black hole case $\kappa$ is the surface gravity. 
We use the absolute value of $t$ to ensure the velocity does not diverge 
for large negative times; however we will be most interested in the 
case $n=2$ so this is moot. 

Asymptotically in time, $\al\to -vn\tau/t^2\to0$, i.e.\ an inertial state, moreover, $\dot z\to0$, so it is not only inertial but asymptotically static, coming to a full stop. 
At $t=0$, the mirror is at the origin and $\dot z=0$ for $n>1$, with 
$\al=0$ if $n>2$ and $\al=2v/\tau$ if $n=2$. 

We will be particularly interested in the case $n=2$, with 
\bea 
PP^2:\ z&=& v\tau\,\ln\left(\frac{t^2}{\tau^2}+1\right)\\ 
\dot z&=&\frac{2v t/\tau}{r^2+1},\quad \al=\frac{v}{\tau}\,\frac{1-r^4}{[(r^2+1)^2-4v^2r^2]^{3/2}} 
\ . \label{accSSD}
\eea 
The maximum velocity equals $v$ for $n=2$.  
We will revisit $PP^2$ as a ``unmoved moving mirror'', which starts at a spatial infinite coordinate at negative infinite time, travels to $z=0$, and then returns to its starting infinity at positive infinite time, being moreover asymptotically static.

\subsection{General Formulas}
The following formulas are used throughout and depend on the trajectory $z(t)$, relevant for emission to one side of the mirror only, i.e.\ the observer's side.  To calculate the energy flux, we use the time dependent version of the Schwarzian derivative without resort to null coordinates (see e.g.\ \cite{Davies:1977yv} or \cite{Good:2013lca}),
\be F(t) = \frac{1}{12\pi}\,\frac{\dddot{z}(\dot{z}^2-1)-3\dot{z}\ddot{z}^2}{(\dot{z}-1)^4(\dot{z}+1)^2} \label{FfromZ}\ . \ee
To calculate total energy emitted, we integrate the energy flux over $u$, where $du = [1-\dot{z}(t)]\,dt$, (see e.g.\ \cite{Good:2016atu} or \cite{Walker:1984vj}),
\be E =\label{EfromF} \int_{-\infty}^{\infty} F(t) (1-\dot{z}(t)) \;d t \ .\ee
To calculate the particle flux, we use the asymptotically inertial character to write the beta Bogolyubov integral in terms of $z(t)$ rather than the shock function $p(u)$:
\be
\label{BfromZ}
\beta_{\w\w'} =
\frac{1}{4\pi\sqrt{\w\w'}}\int_{-\infty}^{\infty} \d t \;
e^{-i\w_{p}t + i\w_{n} z(t)} \left[\w_{p}\dot{z}(t)-\w_{n}\right].
\ee
Here $\w_p \equiv \w + \w'$ and $\w_{n} \equiv \w-\w'$, with $\w$ and $\w'$ the `out' and `in' mode frequencies.  More details on the origin of this formula may be found in \cite{Davies:1976hi}, \cite{Davies:1977yv} or \cite{Good:2016atu}, for example, and references therein.

\section{Self-Dual Mirror Solutions} \label{sec:solutions} 

The moving mirror model has an exact correspondence to black hole collapse for all times, not just late times \cite{Good:2016oey}.  However, the evaporation process in the case of always moving mirrors continues on forever, releasing an infinite number of particles.  As it is known that asymptotically static cases release a finite number of particles \cite{walkerdavies} \cite{Good:2013lca}, we will now examine the particle production of a new asymptotically static solution with time-reversal symmetry.  We also investigate a second asymptotically inertial trajectory which is related but asymptotically drifting rather than static, contrasting their properties with each other and previously known solutions. 


  \begin{figure}[h]
\centering
\mbox{\subfigure{\includegraphics[width=1.6in]{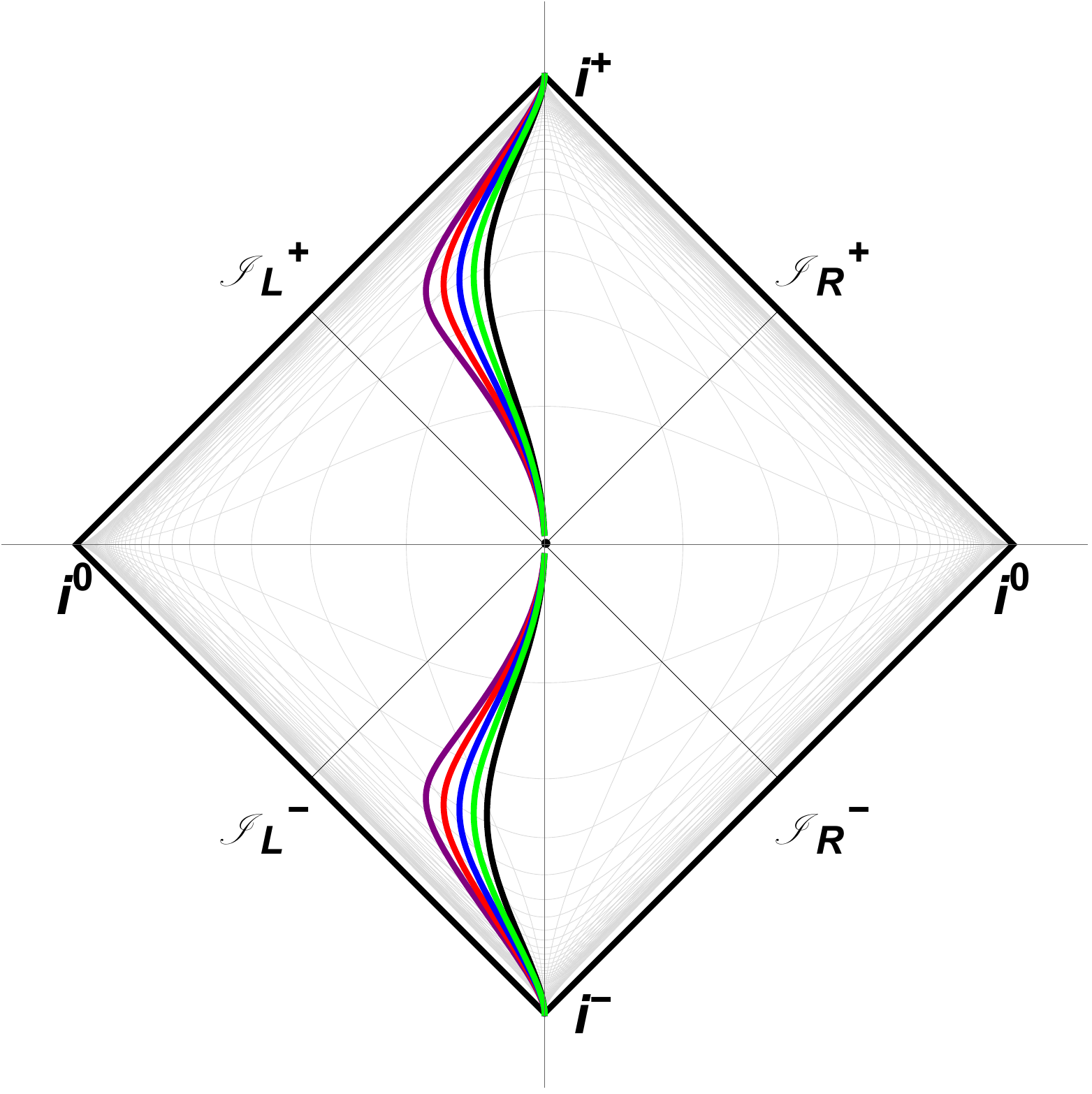}}\quad
\subfigure{\rotatebox{90}{\includegraphics[width=1.6in]{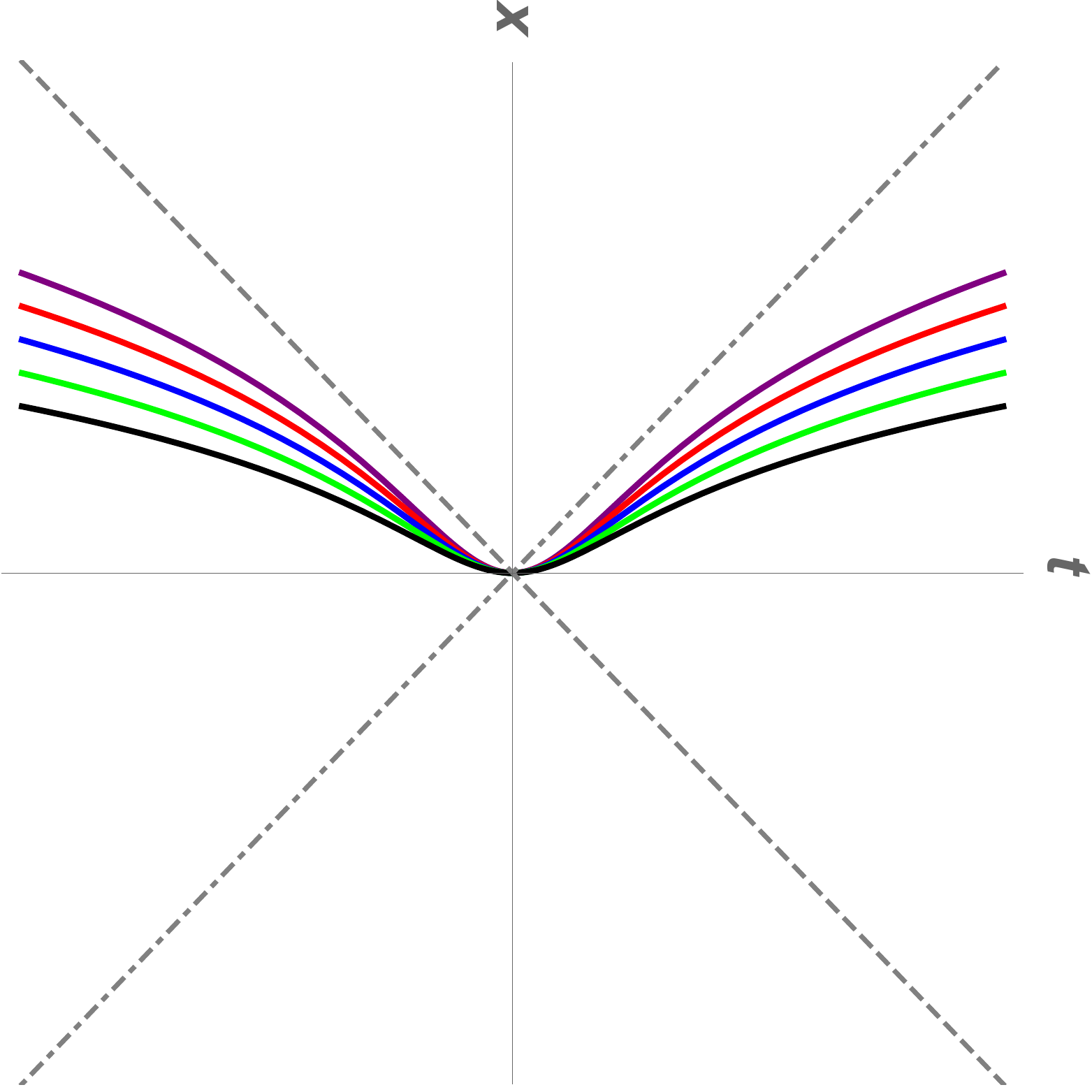} }}}
\caption{\textbf{Left:} In this Penrose diagram, the curves are asymptotically static trajectories, Eq.~(\ref{Zstatic}).  The different maximum speeds correspond to $v =  0.5,0.6, 0.7, 0.8, 0.9$, for black, green, blue, red, purple, respectively. \textbf{Right:} The asymptotically static trajectories are  displayed in the usual spacetime diagram.  The dashed lines represent the light cone.  The trajectories plotted are the same as in the conformal diagram. \label{fig:Penrose}} 
\end{figure}

\subsection{$PP^2$: Self-Dual Moving Mirror -- Asymptotically Static} 

Motivated by symmetry and elevating the energy and particle emission to a more equal footing, it is interesting to examine an exactly solvable moving mirror that is not only finite in total particles or energy, but is self-dual.  A self-dual mirror is one whose particle spectrum is the same on both sides of the mirror. This results in a total energy emission which is the same for both left and right observers. The geometric requirement for self-duality in the moving mirror model is that a trajectory must have zero total displacement, i.e.\ it must be an ``unmoved moving mirror'' (mathematically, the position $z(t)$ should be an even function of time). 

\subsubsection{Trajectory (Static-Self-Dual)}
Consider the trajectory $PP^2$ from the power-power family:
\be z(t) = - \frac{v}{\kappa} \ln (\kappa^2 t^2+1), \label{Zstatic}\ee
where $v$ is the maximum speed of the mirror, $0<v<1$, and $\tau>0$ with $\kappa \equiv \tau^{-1}$. The mirror has its fastest speed at time $t=\tau$.  The mirror starts asymptotically static at infinity, approaches the origin, then returns to its starting infinity, and thus ultimately has zero total shift. It is plotted in Fig.~(\ref{fig:Penrose}). 

\subsubsection{Energy (Static-Self-Dual)}

\begin{figure}[h]
\centering
\includegraphics[width=3.0in]{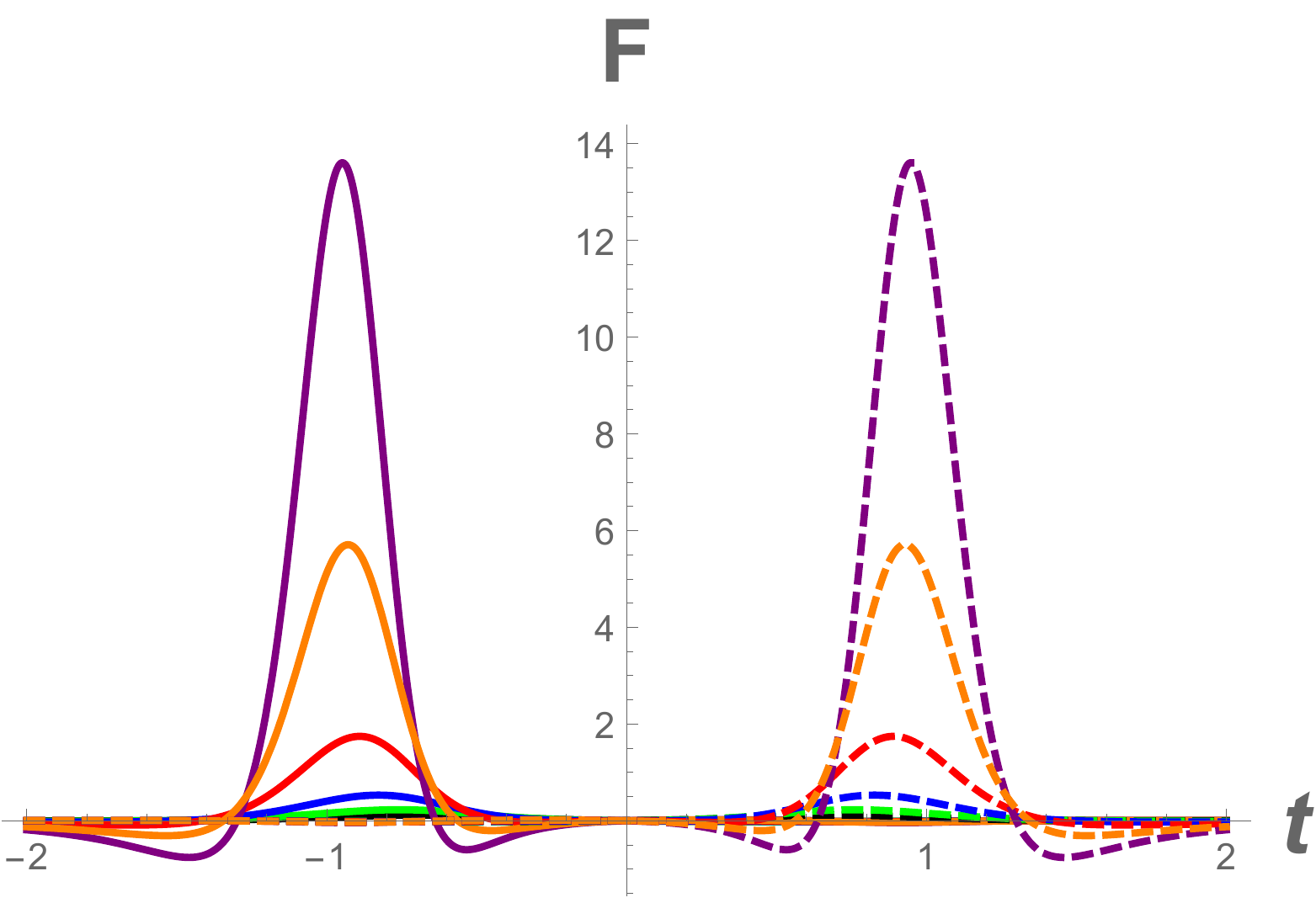}
\caption{The energy flux to the right, $F_R$, and left, $F_L$, of the asymptotically static self-dual mirror, as a function of time in units of $\tau$. The different drift speeds correspond to $v =  0.5, 0.6, 0.7, 0.8, 0.9$, for black, green, blue, red, purple, respectively. Orange is $v=\sqrt{3}/2=0.866.$ The solid (dashed) lines correspond to the energy flux $F_R$ ($F_L$) emitted from the right (left) side, with self-duality giving $F_R(t)=F_L(-t)$. \label{fig:static_energyflux}} 
\end{figure}

 The energy flux from the stress tensor is easily solved for either side (and has two valleys of negative energy flux), with $F_L(t)=F_R(-t)$. The observer on the right sees, using Eq.~(\ref{Zstatic}) in Eq.~(\ref{FfromZ}),
\be F_R(t) = \frac{t \tau  v \left(t^2+\tau ^2\right) \left(3 a \tau ^6-5 \tau ^4 t^2+a \tau ^2 t^4+t^6\right)}{3 \pi  \left(t^2+\tau ^2-2 \tau  t v\right)^2 \left(t^2+\tau ^2+2 \tau  t v\right)^4},\label{SSDflux}\ee
where $a\equiv 2v^2-1$.  Note that the energy flux dies off as $t^{-3}$ for $|t|\gg\tau$. Both $F_L(t)$ and $F_R(t)$ are plotted in Fig.~(\ref{fig:static_energyflux}).

  The total energy emitted to one side is then computed by plugging Eq.~(\ref{SSDflux}) into Eq.~(\ref{EfromF}): 
\be E \equiv E_L = E_R = \frac{(\gamma^2 -1)\left(\gamma ^2+3\right)}{48 \gamma  \tau }, \label{EnergyPP2} \ee
where $\gamma \equiv (1-v^2)^{-1/2}$.  One  immediately sees that if the mirror's maximum speed is zero, $\gamma = 1$, then energy emission is zero, as it should be for a stationary mirror.  The total energy diverges as the mirror's maximum speed approaches the speed of light, $E\to\infty$ as $\gamma\to \infty$. The energy, Eq.~(\ref{EnergyPP2}), as a function of maximum speed, $E(v)$, is plotted in Fig.~(\ref{fig:total_energy}).

\subsubsection{Particles (Static-Self-Dual)}
\begin{figure}[h]
\centering
\includegraphics[width=3.2in]{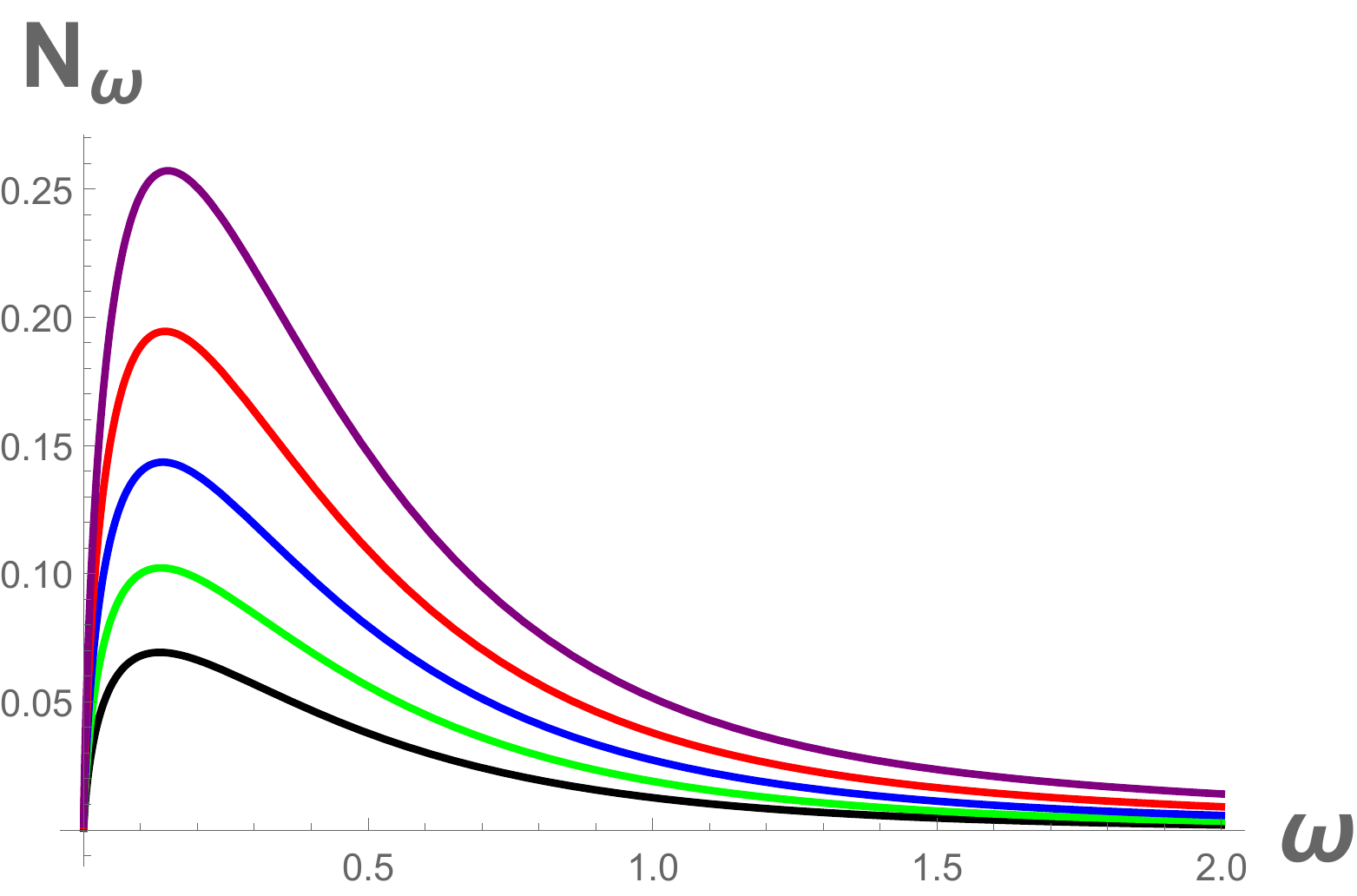}
\caption{The spectrum, $N_\omega$, or particle count per mode, of the asymptotically static self dual mirror is shown above, for the frequency 
in units of $\kappa$.  The same spectrum is obtained for both sides of the mirror, making the mirror self-dual. The different colors, as in Fig.~(\ref{fig:Penrose}), correspond to different maximum speed trajectories: $v =  0.5,0.6, 0.7, 0.8, 0.9$, for black, green, blue, red, purple, respectively. For low speeds, $v\to 0$, the average frequency $E/N = \int \omega N_\omega d\omega / \int N_\omega d\omega \equiv \langle w \rangle \to \kappa/2$.  The peak $N_\omega$ occurs at frequency $\omega_m \approx 0.1$. \label{fig:spectrum}} 
\end{figure}

The beta Bogolyubov coefficient can be found using the trajectory Eq.~(\ref{Zstatic}) in the integral Eq.~(\ref{BfromZ}). The modulus squared is
\be  |\beta_{\omega\omega'}|^2 = \frac{8 v \tau^2  \omega  \omega'}{\pi ^2 \omega_n \omega_p} \sinh (\pi  v \tau \omega_n) \left| K_B(\tau \omega_p)\right|^2, \label{BetaPP2}\ee
where $K$ is the modified Bessel function of the second kind and 
$B\equiv {i v \tau \omega_n +\frac{1}{2}}$.  Here $\omega_p \equiv \omega+\omega'$ and $\omega_n \equiv \omega - \omega'$. The particle spectrum, 
\be N_\omega = \int_0^\infty |\beta_{ \omega  \omega'}|^2 d\omega', \label{spec} \ee
is the same on both sides of the mirror and is plotted in Fig.~(\ref{fig:spectrum}).  

We confirm that associating energy quanta with the particle spectrum reveals consistency between the total energy from the stress tensor, Eq.~(\ref{EnergyPP2}), and the total energy from the Bogolyubov coefficient, Eq.~(\ref{BetaPP2}), 
\be E = \int_0^\infty \int_0^\infty  \omega \, |\beta_{ \omega  \omega' }|^2 \; d \omega \; d \omega'. \ee
The particle count 
\be N = \int_0^\infty \int_0^\infty  |\beta_{ \omega  \omega'}|^2 \; d \omega \; d \omega', \label{NofV}\ee
is finite and easy to numerically solve for any particular maximum speed $v$ of the mirror.  The particle count $N(v)$ is plotted in 
Fig.~(\ref{fig:total_particles}). Both the left and right energies, and left and right particle spectrum densities, $n(\omega,\omega')\equiv|\beta_{\omega\omega'}|^2$, are the same on both sides of the mirror.  This is unlike ``moved moving mirrors'' or ``shifted''  asymptotically static mirrors which have different left-right energies\footnote{The particle count is always the same on both sides of any mirror.  Heuristically, we envision the mirror like a knife slicing vacuum particle pairs apart.  } and left-right spectra (e.g.\ the two previously mentioned known solutions Walker-Davies and Arctx). This self-dual mirror, Eq.~(\ref{Zstatic}), has the most physically left-right symmetric radiation for any moving mirror that has so far been solved.

\subsection{Self-Dual Moving Mirror -- Double Asymptotic Drift}

  \begin{figure}[h]
\centering
\mbox{\subfigure{\includegraphics[width=1.6in]{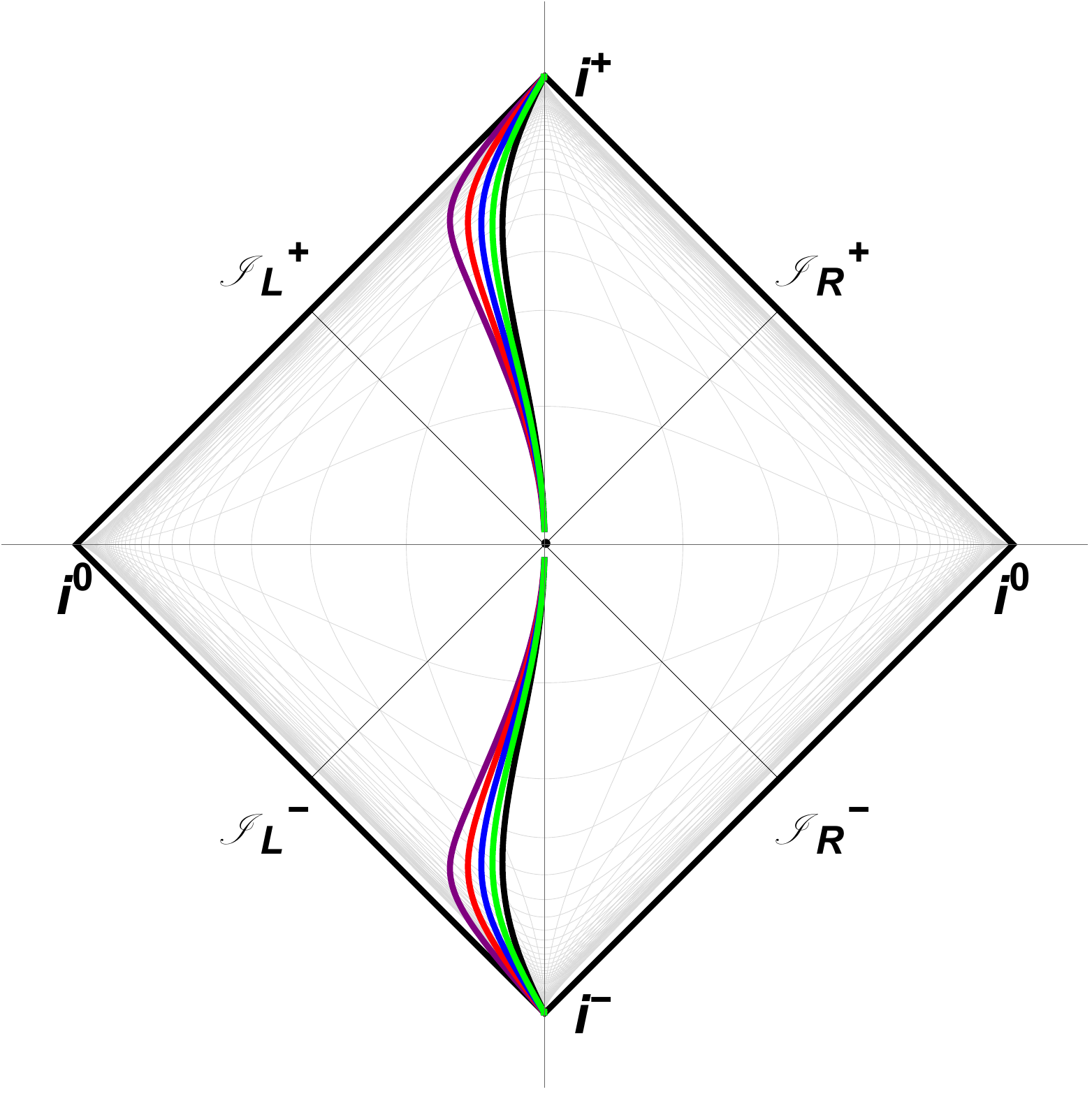}}\quad
\subfigure{\rotatebox{90}{\includegraphics[width=1.6in]{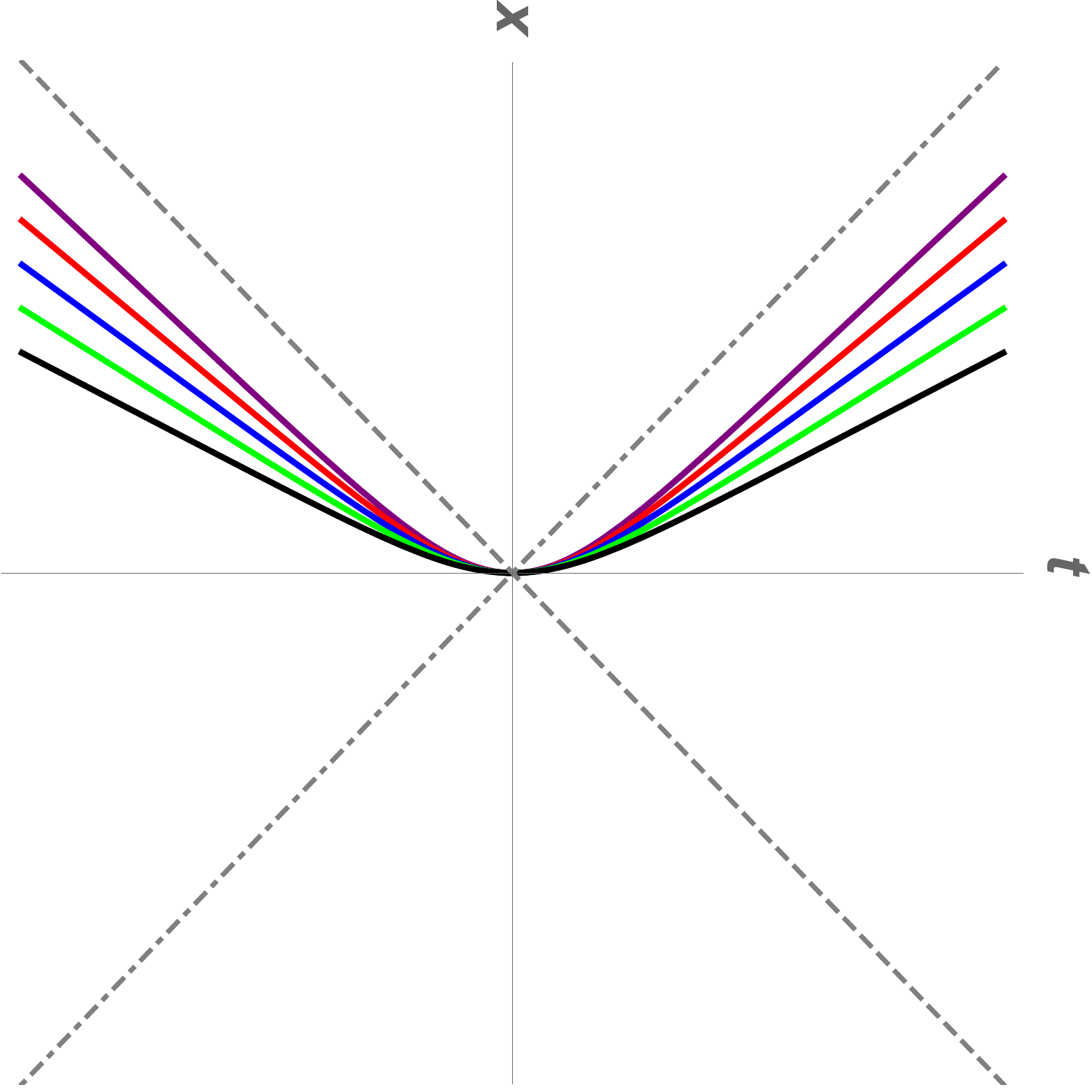} }}}
\caption{\textbf{Left:} In this Penrose diagram, the curves are asymptotically drifting trajectories, Eq.~(\ref{Zdrift}).  The different drift speeds correspond to $v =  0.5,0.6, 0.7, 0.8, 0.9$, for black, green, blue, red, purple, respectively. \textbf{Right:} The asymptotically drifting trajectories with the same coasting speeds are displayed in the usual spacetime diagram.  The dashed lines represent the light cone.  The trajectories here are the same as in the conformal diagram. \label{fig:dd_Penrose}} 
\end{figure}

\subsubsection{Trajectory (Drift-Self-Dual)}
Any of the mirrors in the power-power family are asymptotically static, 
including for arbitrarily large $n$. Moreover, those with $n$ even are self-dual. 
We can generalize Eq.~(\ref{Zstatic}) to 
\be 
z=-v\tau\,\ln(1+c_2 r^2 +c_4 r^4+\dots) \ , 
\ee 
where $r\equiv t/\tau$. For a finite number of terms all of these will be self-dual and asymptotically static. Suppose we now take the infinite series and choose 
the coefficients such that 
\bea 
\tanh:\ z&=&-\frac{v}{\kappa}\,\ln[\cosh \kappa t]\label{Zdrift}\\ 
\dot z&=&-v\tanh \kappa t \ . 
\eea 
Then we get a qualitative change as the mirror is no longer asymptotically static, but asymptotically 
drifts at maximum velocity $v$. The acceleration is 
\be 
\al=\frac{-v\kappa}{[1-v^2\tanh^2 \kappa t]^{3/2}\cosh^2 \kappa t}\ .
\ee 
and so it is still asymptotically inertial, as $\alpha \to 0$ when $t\to \pm \infty$. At $t=0$, $\dot z=0$ while 
$\al=-v\kappa$. Note that for $v=1$, the acceleration takes the particular simple form 
\be 
\al=-\kappa\, \cosh \kappa t \quad (v=1) \ , 
\ee 
and is no longer asymptotically inertial. That is, the tanh model uses a 
multiplicative shift to regularize the acceleration. 

The discontinuity between the asymptotically static nature for a finite 
series and the asymptotic drift nature for the infinite series makes 
it interesting to consider this solution, and the related physics. 
This ``tanh" asymptotically drift mirror, Eq.~(\ref{Zdrift}), has symmetric spectrum and energy flux on both sides also making it self-dual.  However, while the total amount of emitted energy is finite, the total number of particles produced is infinite due to the drifts, see  Fig.~(\ref{fig:drift_spectrum}).

\subsubsection{Energy (Drift-Self-Dual)}
\begin{figure}[h]
\centering
\includegraphics[width=3.0in]{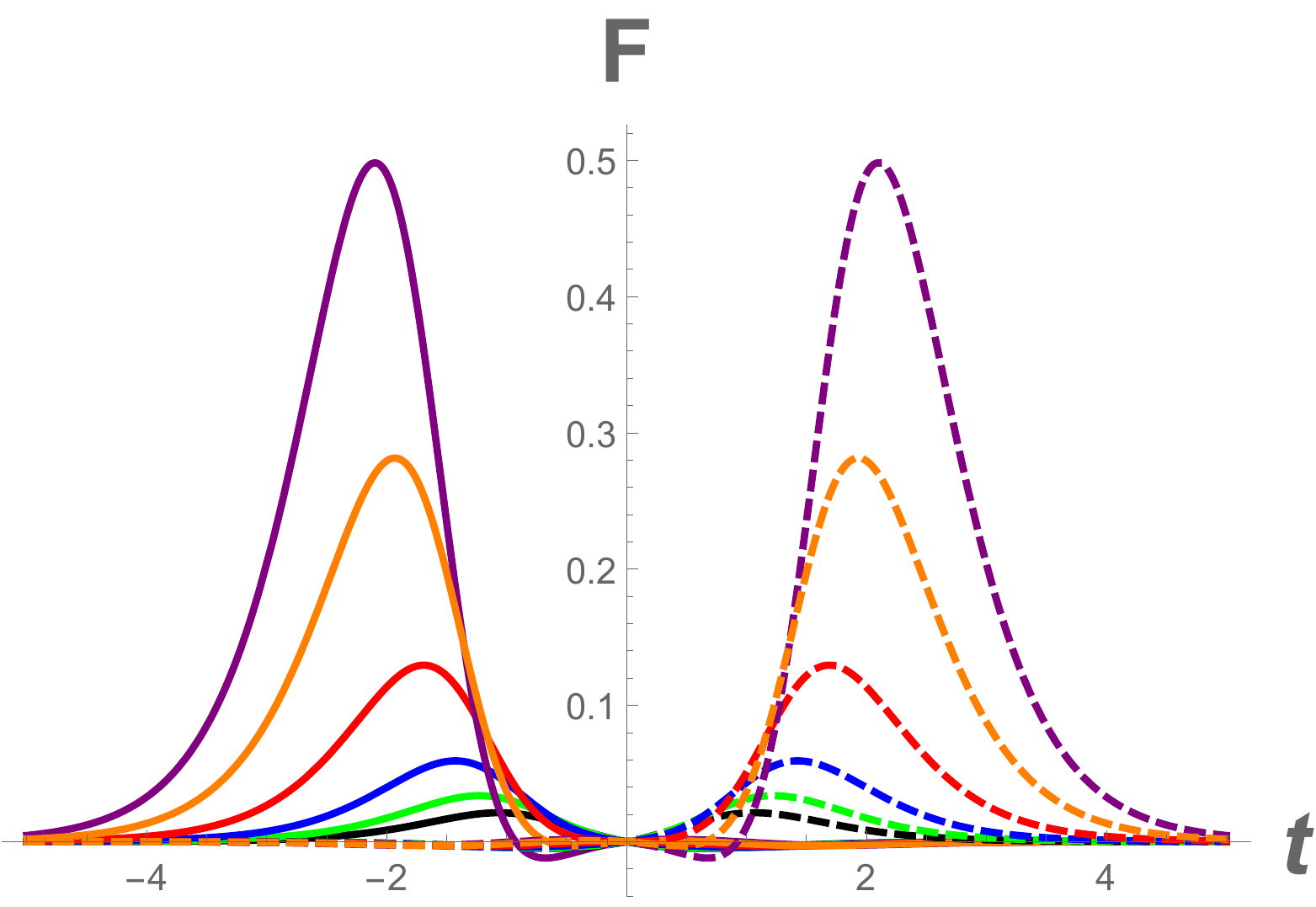}
\caption{The energy flux to the right, $F_R$, and left, $F_L$, of the double asymptotically drift mirror. The different drift speeds correspond to $v =  0.5, 0.6, 0.7, 0.8, 0.9$, for black, green, blue, red, purple, respectively. Orange is $v=\sqrt{3}/2=0.866.$ The solid (dashed) lines correspond to the energy flux emitted from the right (left) side. 
\label{fig:dd_energyflux}} 
\end{figure}

Consider the time-dependent energy flux from the stress tensor, Eq.~(\ref{FfromZ}), emitted by the mirror, Eq.~(\ref{Zdrift}), to the right side, 
\be F(t) = \frac{\kappa ^2}{12\pi}\frac{ A\; \text{sech}^4(\kappa  t) \left[\left(v^2-1\right) \cosh (2 \kappa  t)+2 v^2-1\right]}{ (A-1)^2 (A+1)^4},\label{Fdrift}\ee
where $A\equiv v \tanh\kappa t$.  The energy flux contains two pulses of negative energy flux at sufficiently high speeds, which contrasts with the single pulse of negative energy flux for the drifting mirror in \cite{Good:2016atu} or the drifting mirror in \cite{Good:2015nja}.  The flux, Eq.~(\ref{Fdrift}), is plotted in Fig.~(\ref{fig:dd_energyflux}).

  The total energy emitted by the mirror to one side is easy to find using Eq.~(\ref{Fdrift}) in Eq.~(\ref{EfromF}), 
\be E\equiv E_L = E_R = \frac{2\gamma^2 - 3}{48\pi\tau} + \frac{4\gamma^2-3}{48\pi\tau\gamma^2}\frac{\eta}{v}, \label{Edrift} \ee
where $\eta \equiv \tanh^{-1}v$ is the maximum rapidity, and $\gamma \equiv (1-v^2)^{-1/2}$. This result is plotted in Fig.~(\ref{fig:total_energy}).

\subsubsection{Particles (Drift-Self-Dual)} 
\begin{figure}[h]
\centering
\includegraphics[width=3.2in]{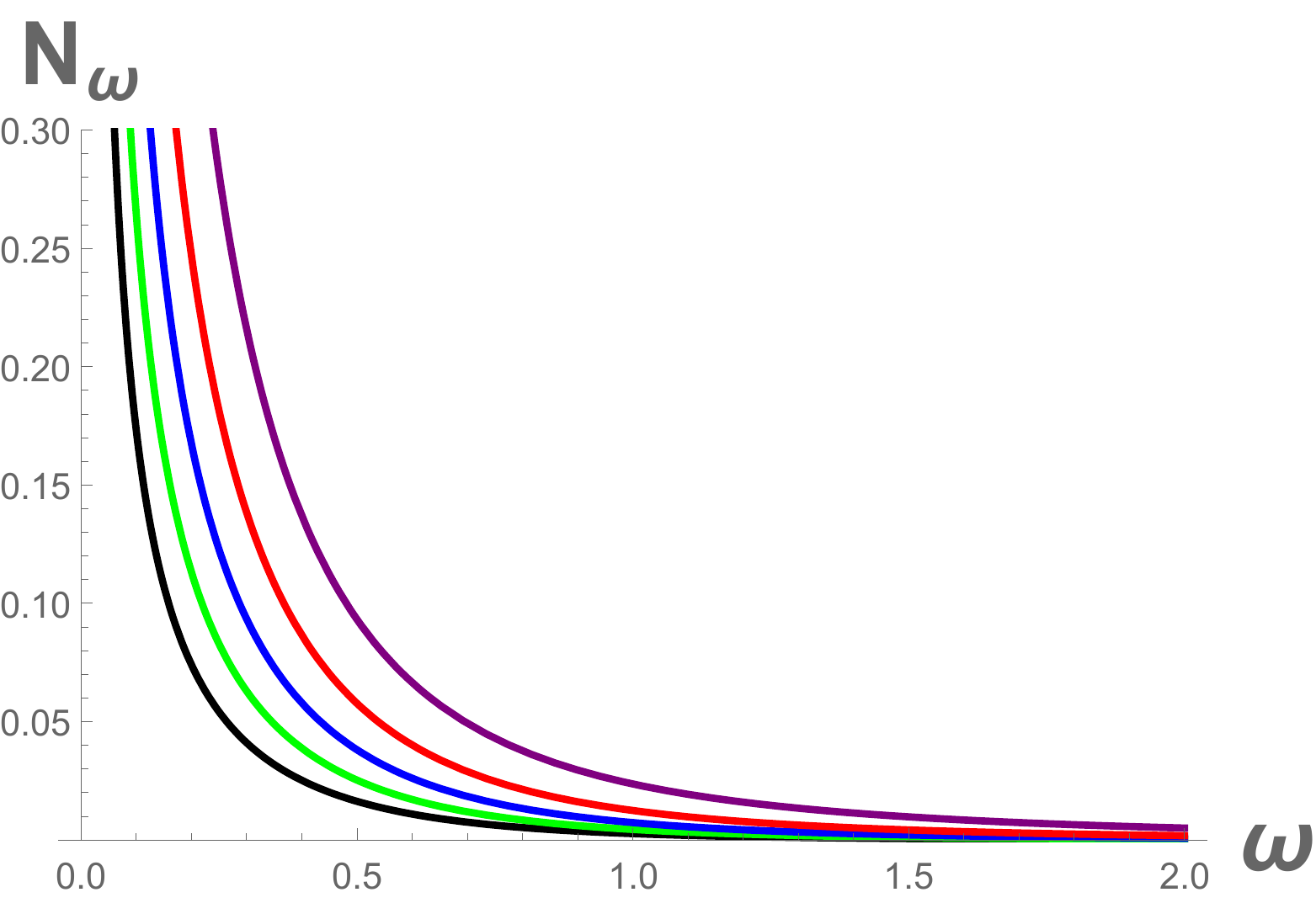}
\caption{The spectrum, $N_\omega$, or particle count per mode, of the asymptotically drifting self dual mirror is shown above.  The same spectrum is obtained for both sides of the mirror, making the mirror self-dual like before. The different colors, as in Fig.~(\ref{fig:dd_Penrose}), correspond to different maximum drift velocities: $v =  0.5, 0.6, 0.7, 0.8, 0.9$, for black, green, blue, red, purple, respectively. Note the increase in 
soft quanta, $\w\to0$. 
\label{fig:drift_spectrum}} 
\end{figure}

The particle emission is found by the beta Bogolyubov coefficient, Eq.~(\ref{BfromZ}), using Eq.~(\ref{Zdrift}) to give 
\be \beta^R_{\omega\omega'} = \frac{2^{i v \tau \omega_n}\sqrt{\omega\omega'}}{2\pi \kappa \omega_n }B(g_-,g_+),\ee
where we have utilized the Beta function (Euler integral of the first kind), $B(x,y) = \frac{\Gamma(x)\Gamma(y)}{\Gamma(x+y)}$. Here $g_\pm \equiv \frac{i}{2}(v\omega_n\pm\omega_p)\tau$. Again, $\kappa\tau=1$, $\omega_p \equiv \omega+\omega'$ and $\omega_n \equiv \omega - \omega'$. The particle spectrum density $n(\omega,\omega') \equiv |\beta_{\omega,\omega'}|^2$ is 
\be n(\omega,\omega') = \frac{v \tau }{2 \pi \omega_n } \frac{\left(\omega_p^2-\omega_n^2\right)}{\left(v^2 \omega_n^2-\omega_p^2\right)} \frac{ \sinh (\pi  v \tau \omega_p)}{ \cosh (\pi  v \tau\omega_n)-\cosh (\pi \tau \omega_p)}. \ee
To find the spectrum, $N_\omega$, one calculates Eq.~(\ref{spec}), which is plotted in Fig.~(\ref{fig:drift_spectrum}).  

It has symmetry, where $n\equiv n(\omega,\omega')$, such that
\be 
\int_0^\infty \int_0^\infty  \omega n \; d \omega \; d \omega' = \int_0^\infty \int_0^\infty  \omega' n \; d \omega \; d \omega'.
\ee
One may associate a quantum of energy $\omega$ (or $\omega'$) with each of the particles and still obtain the same total energy, $E=E_R=E_L$. 

\begin{figure}[h]
\centering
\includegraphics[width=3.2in]{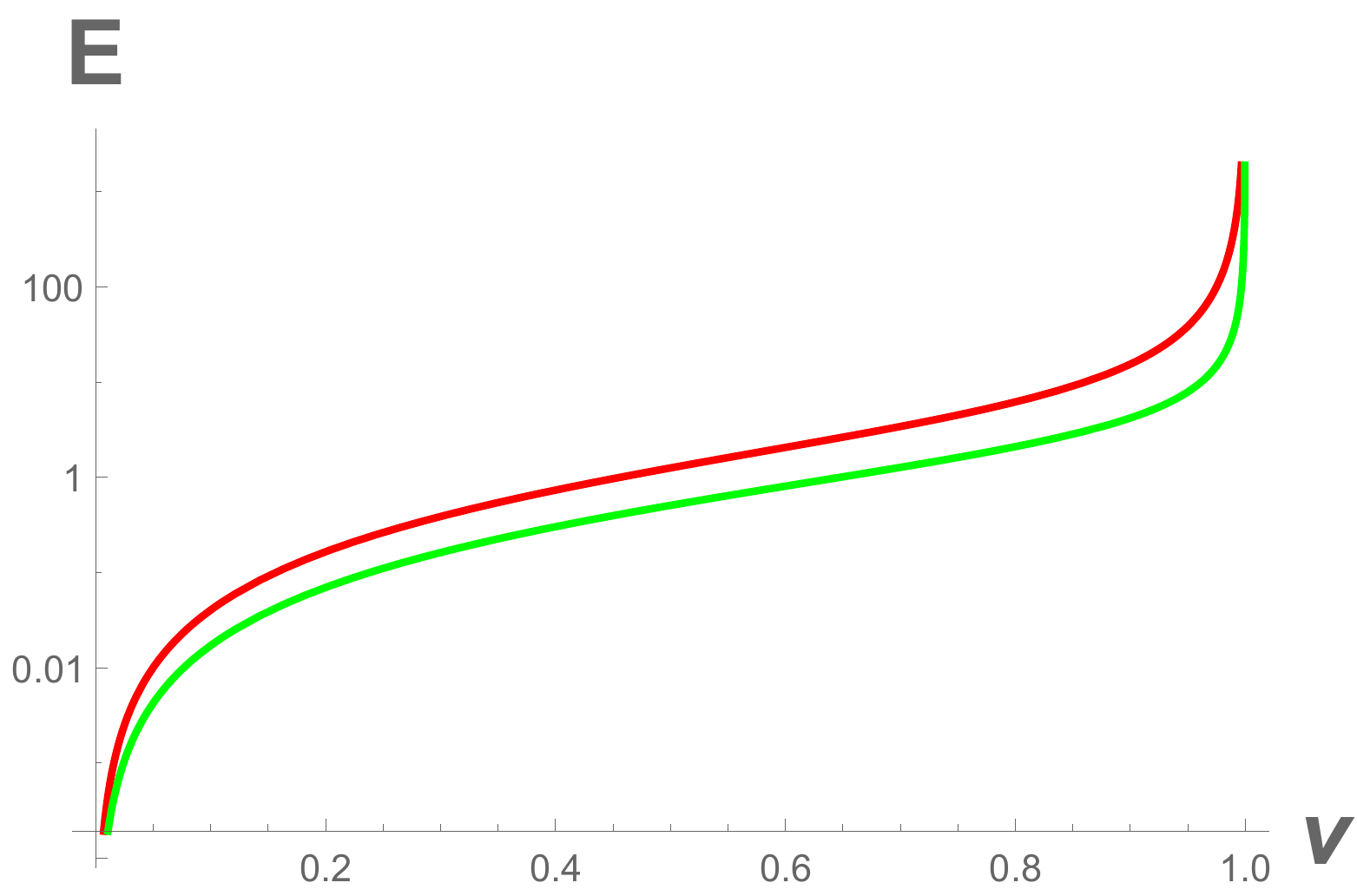}
\caption{A log plot of the energy, in units of $\kappa$, as a function of maximum speed demonstrates that the asymptotically static mirror (red) emits a greater amount of energy than the asymptotically drift mirror (green).   \label{fig:total_energy}} 
\end{figure}

\begin{figure}[h]
\centering
\includegraphics[width=3.2in]{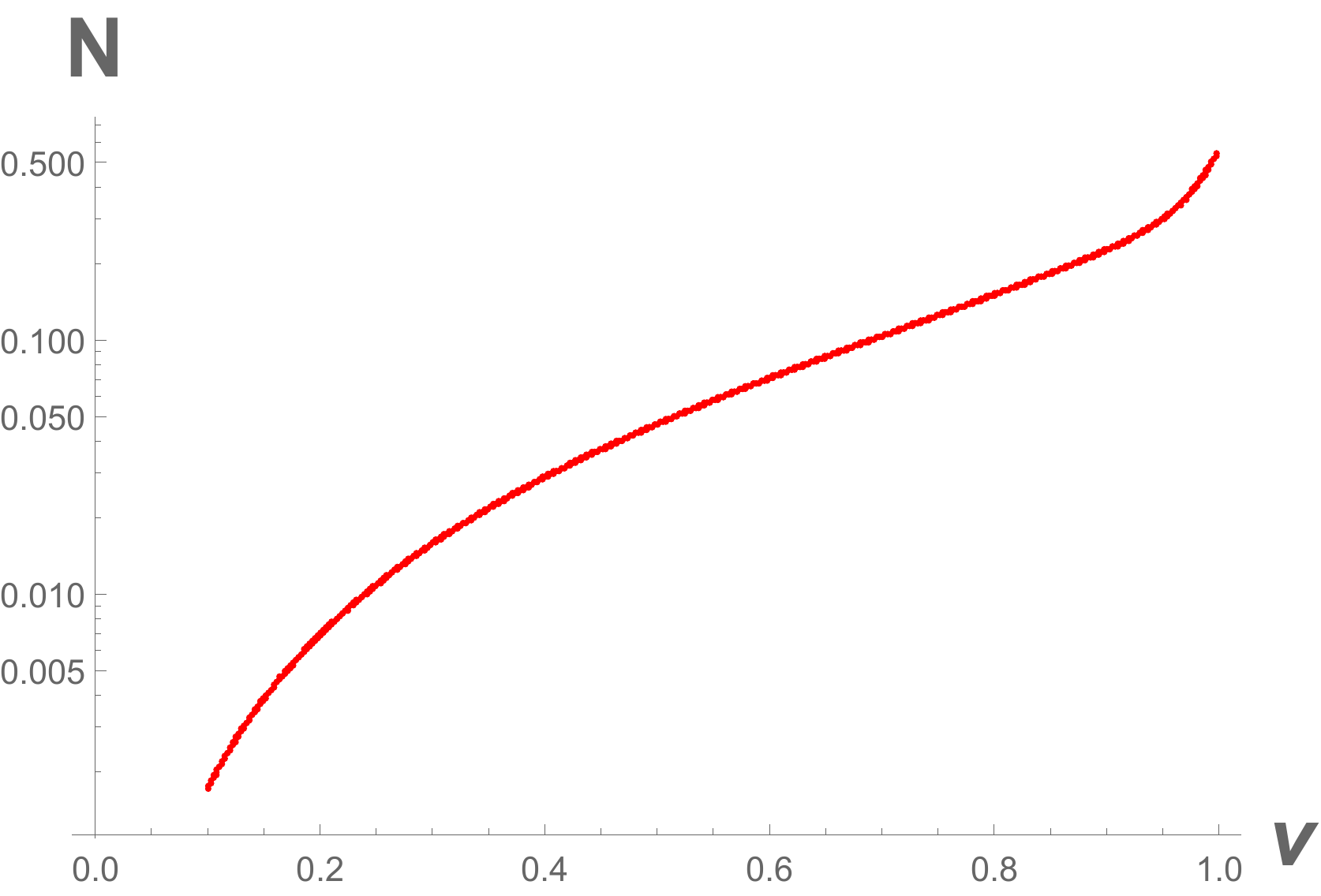}
\caption{A log plot of the particles as a function of maximum speed demonstrates that the static mirror has significantly increased particle count towards the speed of light.  The particle count is independent of $\kappa$ for any given $v$. Here speeds are relativistic beginning with $v=0.1$.  Contrary to what might be anticipated from the energy behavior, the particle count is finite as the speed approaches that of light, $v\to 1$, with $N = 0.5$.  \label{fig:total_particles}} 
\end{figure}

Interestingly, the asymptotically static mirror has greater total energy emission than the asymptotically drift mirror, as seen in Fig.~(\ref{fig:total_energy}). Certainly, one can see from the analytic expressions that the static case has energy flux falling off as $t^{-3}$ while the drift case has exponential die off, as $e^{-2\kappa t}$. As well, for high velocities, the static case total energy scales as $\gamma^3$ while the drift case goes as $\gamma^2$. For low velocities, both scale as $v^2/\tau$, as they must, but the static case has a coefficient of $1/24$ while the drift case has the smaller $1/(9\pi)$. 
Intuitively, one might view this as the asymptotically static mirror taking time to reach speed $v$ from its static state, while the asymptotic drift mirror already starts at speed $v$. 

Regarding total particles emitted, the asymptotically static mirror has a finite number while the asymptotically drift mirror has infinite. Fig.~(\ref{fig:total_particles}) shows the total particle count, $N(v)$, for the asymptotically static mirror, Eq.~(\ref{Zstatic}), as a function of maximum speed, $v<1$.

\section{Entropy and Black Hole Analogs} \label{sec:blackhole}

\subsection{Total Energy from Entropy Flux}

In terms of the trajectory $z(t)$, the von Neumann entanglement entropy flux is \cite{Good:2015nja}
\be S(t) = -\frac{1}{6} \tanh^{-1} \dot{z}(t) = -\frac{1}{6}\eta(t), \ee
where $\eta(t)$ is a time-dependent rapidity (note in $1+1$ dimensional motion, rapidities are additive).    In terms of the ray-tracing function $p(u)$ or $\eta(u)$ the entropy as a function of null coordinate $u=t-x$ is
\be S(u) = -\frac{1}{12} \ln p'(u) = -\frac{1}{6}\eta(u). \ee 

One can express the total finite energy emitted by a mirror to a single observer using
\be E = \frac{1}{12\pi} \int_{-\infty}^\infty \eta'(u)^2 du = \frac{3}{\pi} \int_{-\infty}^\infty S'(u)^2 du , \ee
where the prime is a derivative with respect to $u$.  This is re-expressed in $t$ rather than $u$ as 
\be 
E = \frac{1}{24\pi} \int_{-\infty}^{\infty} dt\,\dot{\eta}(t)^2\,\left[1+e^{2 \eta(t)}\right]\,, \ee
or, as an energy-entropy relationship, 
\be
E=\frac{3}{2\pi} \int_{-\infty}^{\infty} dt\,\dot{S}(t)^2\,\left[1+e^{-12 S(t)}\right]\,. \label{energyfromentropy}\ee
This relation must hold to maintain global consistency between the total energy and the entropy flux, i.e.\ using the total energy derived from the beta Bogolyubov coefficients or from the stress tensor, the correct entropy flux must satisfy Eq.~(\ref{energyfromentropy}).  It holds particular relevance for checking any serious tension or inconsistency between the entanglement entropy and the energy flux \cite{Chen:2017lum}. 

\subsection{Entropy of Self-Dual Solutions}
\begin{figure}[h]
\centering
\includegraphics[width=3.2in]{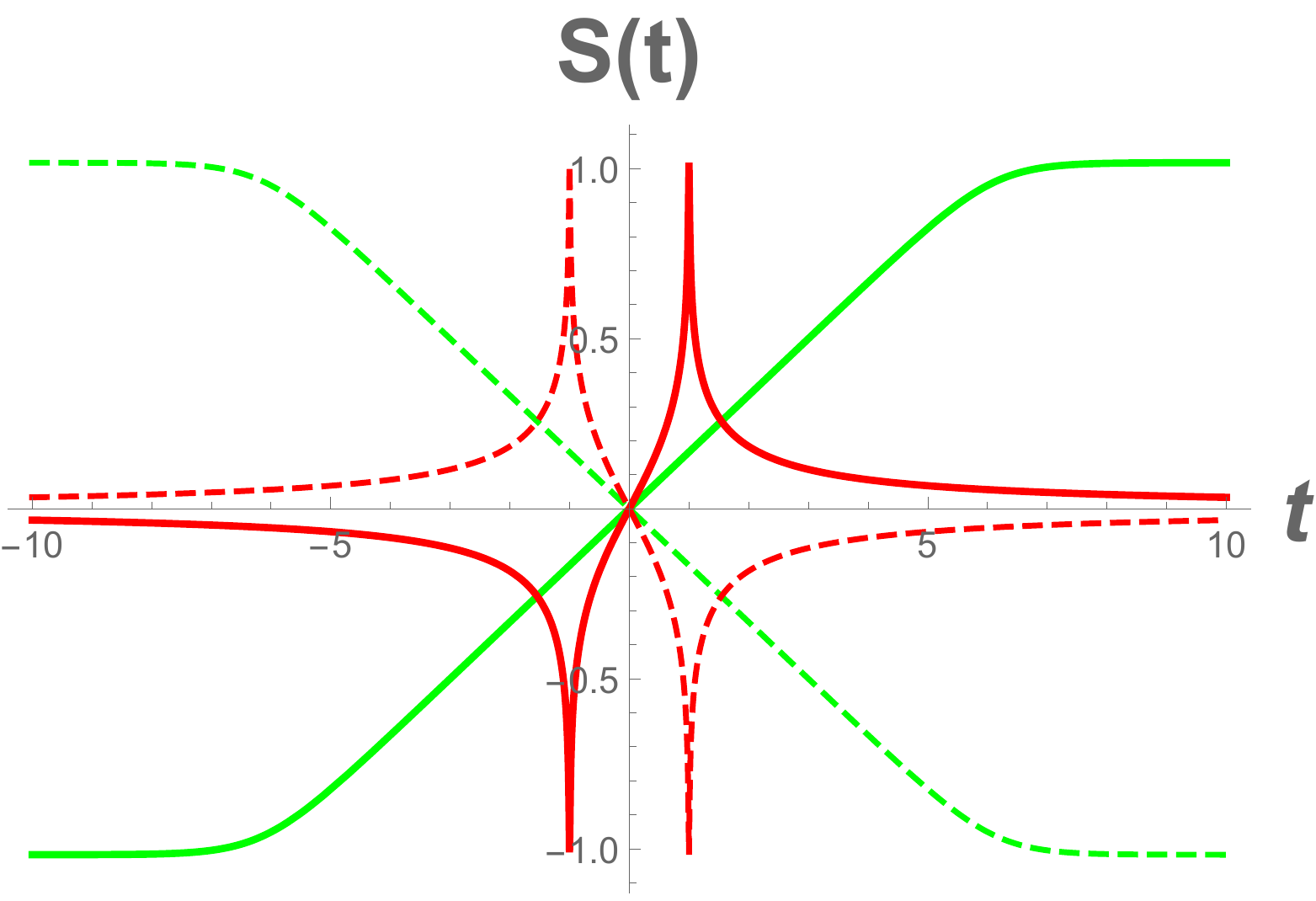}
\caption{The entropy flux, $S(t)$, of the self-dual solutions.  Red (green) indicates the asymptotically static (drifting) mirror.  Dashed (solid) lines indicate the left (right) sides of the mirror. Here time is in units of $\tau$ and $v=0.99999$.  The asymptotically static mirror has a finite total entropy flux, even at high speeds.    \label{fig:entropyflux}} 
\end{figure}

\begin{figure}[h]
\centering
\includegraphics[width=3.2in]{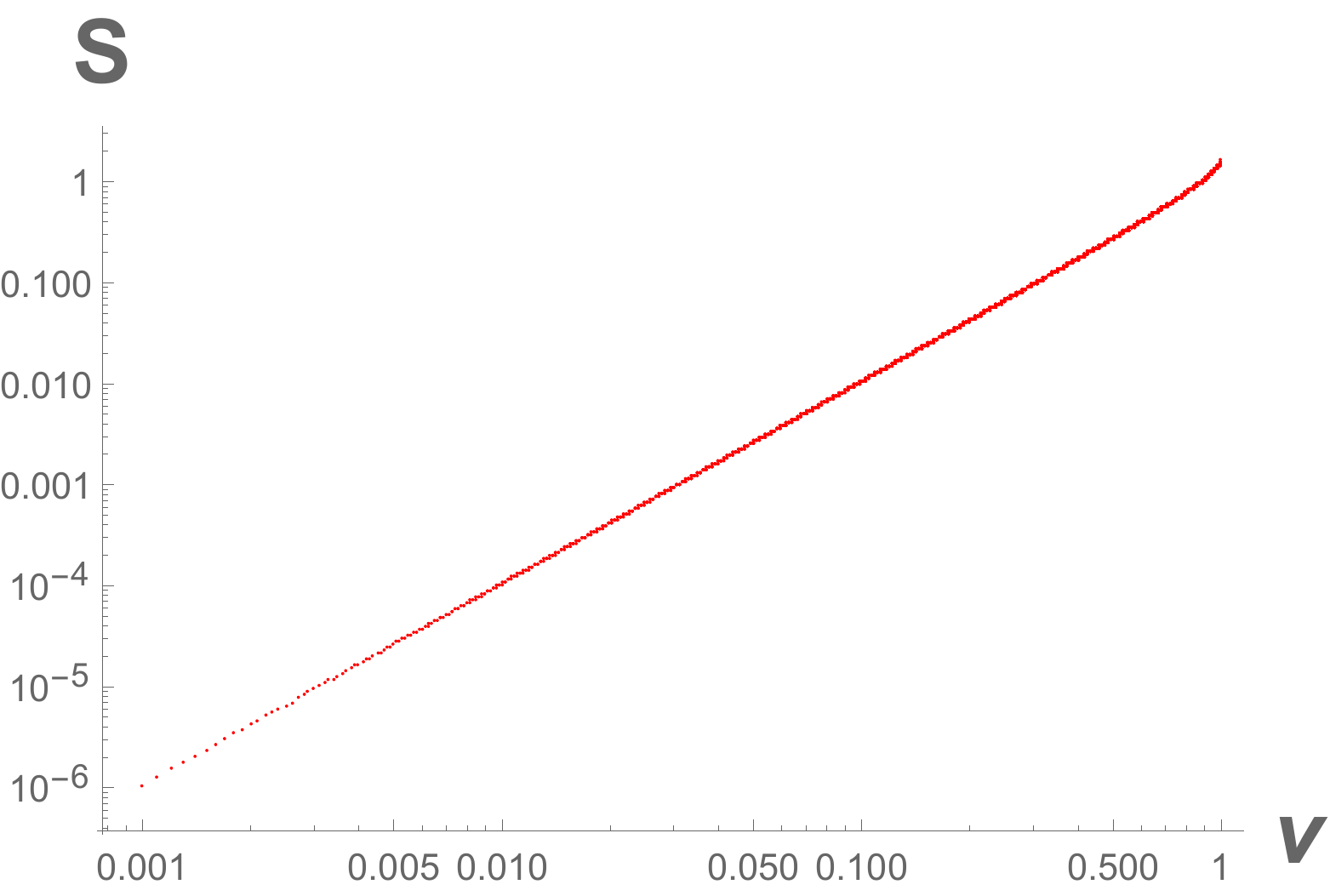}
\caption{The total von Neumann entanglement entropy, in units of $\tau$, as a function of maximum speed, $S(v)$, of the self-dual asymptotically static solution calculated numerically using a double exponential integration method.  Contrary to what might be anticipated from the energy behavior, but consistent with the total finite particle count, the entropy is finite even when the mirror approaches the speed of light, $v \to 1$, with $S \to 1.64$. \label{fig:totalentropy}} 
\end{figure}

The entropy flux for both the self-dual solutions are plotted in Fig.~(\ref{fig:entropyflux}).  The entropy flux emitted to the right side for the asymptotically static and drifting mirrors are respectively,
\bea 
S_R(t)&=& \frac{1}{6} \tanh ^{-1}\left(\frac{2v \kappa  t}{\kappa ^2 t^2+1}\right)\quad ({\rm static})\ ,\label{entf1} \\ 
S_R(t)&=& \frac{1}{6} \tanh ^{-1}(v \tanh \kappa  t) \quad ({\rm drift}) \ .\label{entf2}
\eea 
Time-reversal symmetry gives $S_R(t)=S_L(-t)=-S_L(t)$. We have confirmed the consistency of the entropy flux, Eq.~(\ref{entf1}) and Eq.~(\ref{entf2}), with the total energy emission, Eq.~(\ref{EnergyPP2}) and Eq.~(\ref{Edrift}), through Eq.~(\ref{energyfromentropy}). Notice that the entropy flux is an odd function of $t$, meaning that integration at the mirror gives
\be \mathbb{S} = \int_{-\infty}^{+\infty} S_R(t) dt = 0\ . \ee
The finite total entropy for the asymptotically static case for an observer at $u=\infty$ 
to the right, $\mathscr{I}^+_R$, is 
\be S = \int_{-\infty}^{+\infty} S_R(t) \,[1-\dot{z}(t)]\;dt, \ee
and is plotted in Fig.~(\ref{fig:totalentropy}) as a function of maximum speed, $S(v)$. 

Note the total energy in both self dual cases is simply linearly proportional to $\kappa$, the acceleration parameter, which sets the scale of the problem. The total particle count (of course, a dimensionless quantity) is independent of $\kappa$.  The total entropy is linearly proportional to the characteristic time   $\tau = \kappa^{-1}$. 

The asymptotically drifting mirror has a finite energy emission ($E\propto \kappa$) and so as the mirror stops accelerating, $\alpha(t\to\infty)\to 0$, the mirror stops radiating, $F(t\to\infty) \to 0$.  However, the infinite total particle production is aligned with the feature of ongoing redshift of the field modes.  Even though the radiation has fully stopped at late-times, the drifting itself becomes an essential feature of the late-time physics.  While all mode frequencies are reflected by the mirror, and at late-times there is no particle production, there is still an eternal red-shifting of field modes This has the important implication that this model does not describe complete black hole evaporation: the analog black hole leaves behind a remnant. 

The asymptotically static mirror avoids these complexities. In the black hole analog, particles that travel through empty space at late times encounter no black hole, and no remnant; they reflect off a static mirror, with zero redshift.  This is a simple demonstration case of the preservation of unitarity: any pure initial state evolves smoothly into a pure final state.  Furthermore the entropy production is finite. The asymptotically static mirror models a black hole that evaporates completely.

\section{Conclusions} \label{sec:concl} 

A key result of this work has been the demonstration of a physically simple solution to the moving mirror model by symmetrizing the particle spectrum on both sides of the mirror, through applying time-reversal symmetry on the trajectory motion, $z(t) = z(-t)$. In the asymptotically static case, the results are fully finite physical observables.  Moreover, time-reversal symmetry results in an energy symmetry from the two sides of the mirror, demonstrating the first self-dual solution to the dynamical Casimir effect.
  
Since the particle count is finite and generalized to a two parameter system $(\kappa,v)$, of the acceleration parameter and maximum speed respectively, we were able to carry out the first investigation of total particle count from a relativistically moving mirror as a function of its maximum speed for all speeds less than the speed of light.  As is intuitive, the particle emission is monotonic and increases with maximum speed.  Less intuitively, the particle emission is finite even when the maximum mirror speed $v\to 1$, in contrast to the asymptotic energy divergence as $v\to 1$.   

This calculability led to an investigation into the self-dual asymptotically \textit{drifting} case where infinite particle production exists and is associated with the coasting end-state (black hole remnant analog\footnote{For other moving mirror solutions which act as black hole remnants see e.g.\   \cite{Good:2016atu}\cite{Good:2013lca}\cite{Wilczek:1993jn}\cite{Good:2015nja}.}).  We have identified this as a limit of infinite terms in the trajectory of motion, $z(t)$. 

We then demonstrated consistency between the entropy flux and the total energy radiated for the self-dual solutions.  We found the total finite entropy generated for the asymptotically static case, demonstrating no divergences even for high speeds as $v\to 1$, matching the behavior for the finite particle count in this regime. 

An interesting avenue for investigation of this or some other asymptotically static mirror would be the demonstration of $F(\Delta \mathcal{T}) \to \kappa^2/(48\pi)$ for some finite time $\Delta \mathcal{T}$, indicating constant energy flux (as is done in \cite{Good:2016atu} for an asymptotically coasting trajectory).  Or, likewise, demonstrating  $|\beta_{\omega\omega'}|^2 \propto (e^{2\pi \omega/\kappa} -1)^{-1}$ indicating Planckian distributed particles at temperature $2\pi T = \kappa$, for some finite time $\Delta \mathcal{T}$ at some point during the black hole's lifetime.  If this could be shown, then the asymptotically static case would go even further to substantiate the picture of black hole radiation from early times (collapse) to late times (black hole temperature) to very late times (vanishing black holes). 

In the near future, we plan to investigate asymmetrical asymptotically static mirror solutions that have advantages over the known Arctx and Walker-Davies solutions, and a generalizable particle count in terms of the maximum speed of the mirror, as well as a given $z(t)$ trajectory, for $z(t) \neq z(-t)$. 

Intriguingly, particle production studies from moving mirrors, and the associated insights into black hole particle creation, may not be purely theoretical; see \cite{wilson} and \cite{1512.04064} for different possibilities about testing this in the laboratory.

\acknowledgments 
MG appreciates support from the Julian Schwinger Foundation under Grant 15-07-0000 and the ORAU and Social Policy grants at Nazarbayev University. EL is supported in part by the Energetic Cosmos Laboratory and by 
the U.S.\ Department of Energy, Office of Science, Office of High Energy 
Physics, under Award DE-SC-0007867 and contract no.\ DE-AC02-05CH11231.


\end{document}